\documentclass[12pt]{article}
\usepackage[margin=1in]{geometry}
\usepackage{times}
\usepackage{graphicx}
\usepackage{amsmath, amssymb}
\usepackage{titlesec}
\usepackage[super,sort&compress]{natbib}

\titlespacing*{\section}{0pt}{0em}{0em}
\titlespacing*{\subsection}{0pt}{0em}{-0.5em}

\titleformat{\section}
  {\normalfont\fontsize{14}{16}\bfseries}{\thesection}{1em}{}
\titleformat{\subsection}
  {\normalfont\fontsize{12}{14}\bfseries}{\thesubsection}{1em}{}

\title{\bfseries\Large Directional Random Lasing in Liquid Crystal Infiltrated Metasurfaces\\[-0.5em]}
\author{\normalsize Khoi Anh Pham$^{1}$ and \normalsize Giuseppe Strangi$^{1,2}$\textsuperscript{*}}
\date{
\vspace{-1.5em}
{\normalsize
$^{1}$Department of Physics, Case Western Reserve University, 10600 Euclid Avenue, Cleveland, Ohio. 44106, USA\\
$^{2}$NLHT Labs - Department of Physics, University of Calabria, Rende, Italy.
}
}

\begin{document}
\maketitle
\begin{center}
\vspace{-3em}
\text{* Corresponding email: gxs284@case.edu} 
\end{center}
\vspace{-1em}
\begin{abstract}
\vspace{-1em}
\normalsize   
\noindent
Random lasers (RL) emit light through multiple scattering in disordered gain media, typically resulting in isotropic emission with limited directionality control. Controlling RL emission direction in compact systems remains a challenge. Here we report directional random lasing achieved by infiltrating dye-doped nematic liquid crystals into a nanostructured silica metasurface. By adjusting pump energy, we induce a transition from uniform angular photoluminescence to a strongly directional emission peak at large angles in the amplified spontaneous emission and RL regimes. This directionality arises from enhanced spatial coherence in the strong scattering regime, enabling coupling of guided random-laser modes to high-angle diffraction through the metasurface grating. Our system demonstrates wide-angle RL beam steering at submicron scale without complex external components. These results provide a straightforward method to control RL emission directionality, advancing tunable coherent light sources and metasurface-based photonic applications.
\end{abstract}
\vspace{-1em}
\section*{Introduction}
\vspace{-0.5em}
Random lasers are light sources that provide optical gain through stimulated emission, like conventional lasers, but lack a fixed feedback cavity.  Rather, random laser generation arises from multiple scattering, where interference and phase correlations between scattered photons determine the lasing modes \cite{Cao2003}\cite{Sapienza2019}\cite{Cao2000}. In an amplifying disordered media, like dye-doped liquid crystals (LC), photons would be amplified and scattered multiple times, undergoing a random walk \cite{Ahmad2025}\cite{Wan2017}\cite{Strangi2006}\cite{Shang2022}. Even in highly disordered materials, multiple scattering can preserve partial phase correlations over finite lengths, providing sufficient feedback for lasing oscillations. However, to generate random lasing, the gain of the system must overcome the loss, which is determined by various parameters such as the volume threshold V$_{\text{th}}$, energy threshold E$_{\text{th}}$ for population inversion, and the scattering mean free path $\ell_s$ and transport mean free path $\ell_t$ with respect to the system characteristic length $L$ \cite{Hara2024}\cite{Kumar2023}. Before reaching the random lasing regime, the system enters the amplified spontaneous emission (ASE) regime, showing a stronger intensity and narrower emission bandwidth than photoluminescence (PL) but without the spiking features of lasing \cite{Umar2019}\cite{Wu2008}. However, both ASE and RL lacks directionality, typically producing isotropic emission profile due to the random multiple scattering.
\begin{figure*}[!t]
    \centering
    \includegraphics[width=\linewidth]{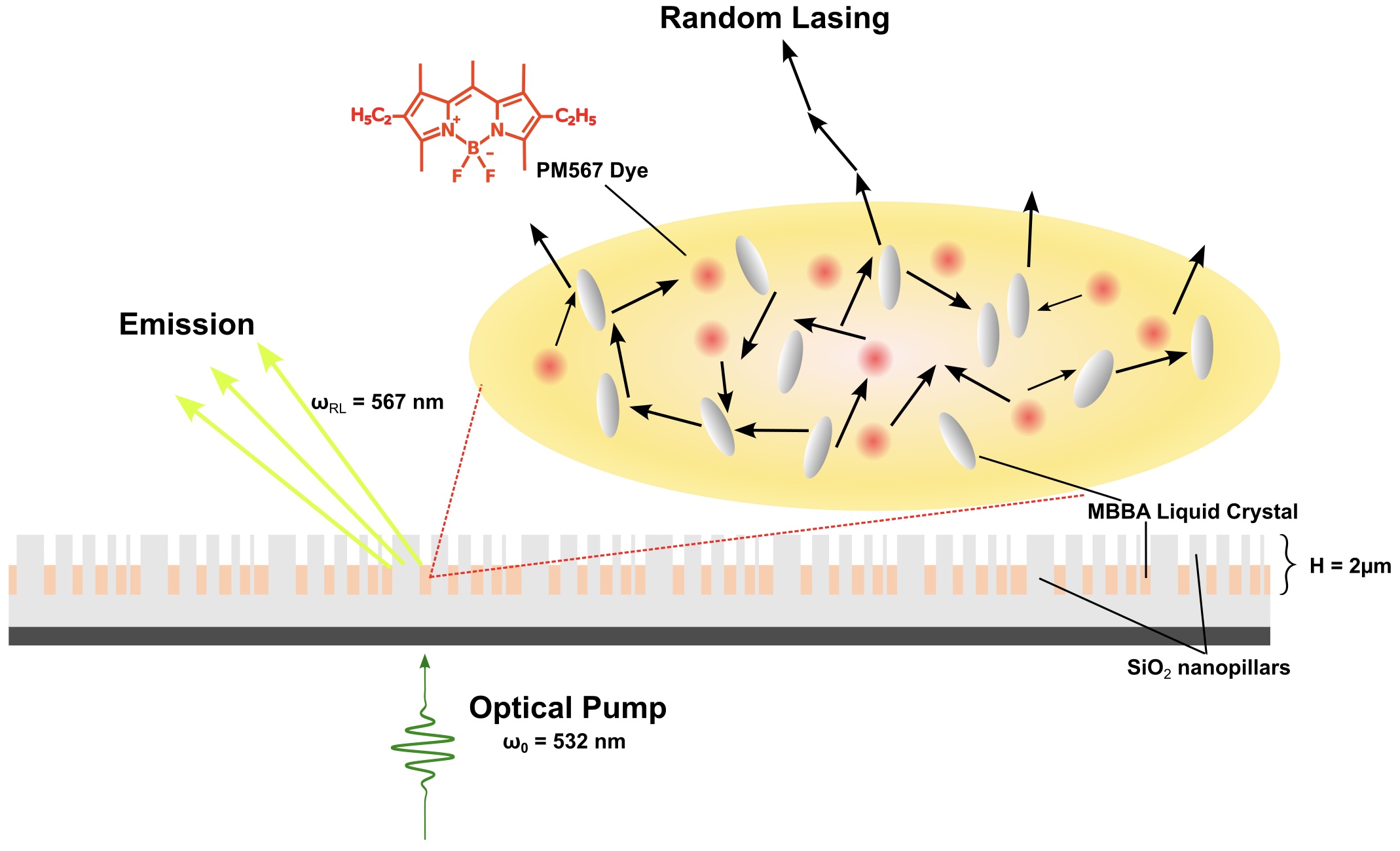}
    \caption{\textbf{Schematic illustration of random lasing in a LC-infiltrated metasurface.} A fused-silica metasurface (area = 0.785 cm$^2$) comprising ~140 million nanopillars is infiltrated with dye-doped LC via hemiwicking. Under optical excitation, the random laser emits directionally at large angles relative to the surface normal.
}
    \label{fig:Fig1}
\end{figure*}
A key objective is to achieve control over the directionality of random lasing. Previous attempts have managed to control RL’s direction through using a soliton that collected and guided the RL, the use of a holey dielectric metal cavity, a self-assembled biological waveguide structure, and a conical structure \cite{Perumbilavil2018}\cite{Schnhuber2016}\cite{Zhang2014}\cite{Wang2025}. These studies employed structured-base methods to guide the RL to emit in a preferred direction, typically along the direction of pumping. However, the largest downside in past attempts is the use of bulky waveguides and thick LC cells for high gain to generate the RL. 

Here, we report directional RL generation using a dye-doped nematic liquid crystal (NLC) infiltrated inside a metasurface (Fig.\ref{fig:Fig1}). Metasurfaces are thin planar array of nanostructured material at the subwavelength scale. Large research efforts have demonstrated metasurfaces’ vast abilities to shape light such as focusing, orbital-angular momentum light generation, and beam steering through controlling the phase, amplitude, polarization of light \cite{Yu2011}\cite{Chang2024}\cite{Zhuang2023}\cite{Zhang2018}\cite{Shalaginov2021}. Importantly, metasurfaces rely on coherent light to efficiently control the phase, and leverage interference for wavefront shaping. Lower coherence source such as PL from laser dye therefore can not be optimally structured using metasurfaces, limiting the possible tasks that can be performed. However, by increasing the energy input or lowering the lasing threshold of the system, dye emitters can gain partial coherence through multiple scattering, allowing integration into metasurfaces. A light-emitting binary metasurfaces with distributed feedback have demonstrated holographic lasing leveraging such partial coherence \cite{Bashiri2025}. Here, we use a metasurface designed with a quadratic phase profile and we demonstrate theoretically and experimentally that the metasurface itself acts as an internal light source, transitioning from isotropic to highly directional emission as the coherence of the random lasing increases, all within a submicron-scale structure.
\section*{Results}
\subsection*{Liquid Crystal Infiltrated Metasurfaces
}
\begin{figure*}[!t]
    \centering
    \includegraphics[width=0.99\linewidth]{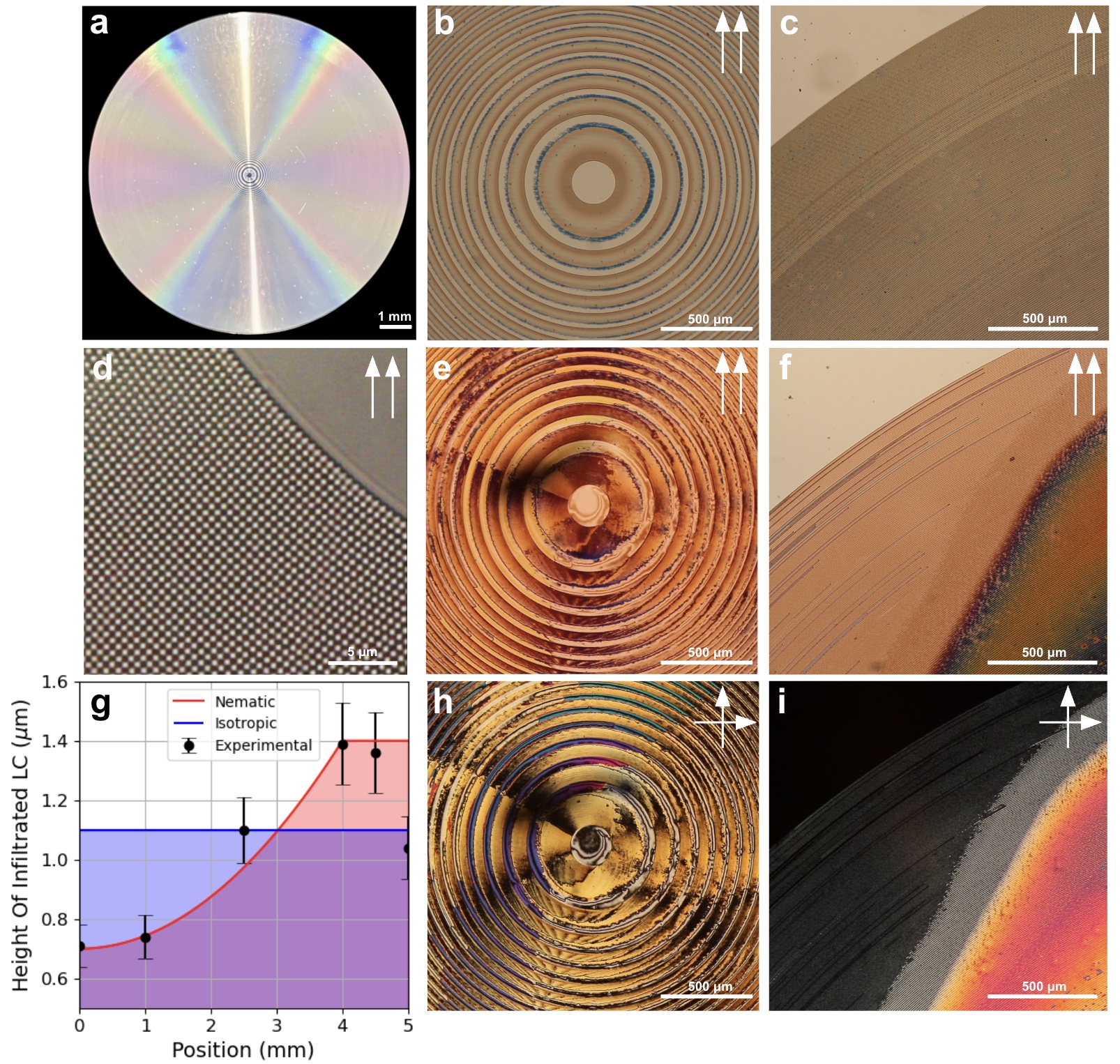}
    \caption{\textbf{Infiltrated SiO2 metasurface.} (a) Full scale fabricated metasurface with 1 cm diameter. (b), (c) Uninfiltrated metasurface under co-polarization with concentric rings at the center (edge). (e), (f) Metasurface infiltrated with dye doped LC at the center (edge) under co-polarization and (h), (i) cross-polarization. (d) Nanopillars of the metasurface taken with a x100 objective near the center of the metasurface. (g) Infiltration height profile as a function of metasurface radius. }
    \label{fig:Fig2}
\end{figure*}
We use all-glass (SiO2) metasurfaces, shown in Fig.\ref{fig:Fig2}a with refractive index $n_{\text{SiO}_2}$ = 1.45, diameter $d$ = 1 cm, height $h$ = 2 $\mathrm{\mathrm{\mu}}$m, and edge-to-edge spacing of 250 nm designed to operate in free space at wavelength $\lambda$ = 633 nm. The widths of the pillars range from 250-600 nm spanning a $\pi$ phase delay range whereas no pillars is considered as zero phase delay. The design thus creates concentric rings of pillars with empty spacing between the rings. These concentric rings create the necessary Fresnel zones (FZ) for light interference. Further information regarding the metasurface fabrication is discussed in detail in \cite{Park2019}. We are using birefringence MBBA rod-shaped NLC (Sigma Aldrich) doped with 1wt\% PM567 laser dye (Luxotica) as our RL system. The NLC provides strong scattering due to the thermal fluctuation in local ordering about the director \cite{Stallinga1996}. The long-range order of the NLC infiltrated inside the metasurface creates an effective change in refractive index for different polarization of emission due to the birefringence. Crucially, the infiltration of NLC ($\Delta n$ = 0.12, $n_e$ = 1.68, and $n_o$ = 1.56 at 567 nm) altered the refractive index contrast with the metasurfaces ($\Delta n$ = - 0.2) compared to being in air ($\Delta n$ = 0.45), effectively reducing the phase range induced by the metasurface. The smaller phase range lower the grating diffraction efficiency due to a weaker interference effect. However, given the relatively large normalized unit cell period P/$\lambda$ = 0.8-1.4 of the metasurface (where P is the period of a unit cell), the grating efficiency offered can still be high due to the suppressed higher-order guided Bloch modes in low-index metasurface \cite{Sacchi2025}. Furthermore, the infiltration of the NLC opens the possibility of a waveguide mode due to $n_{\text{air}}$ $<$ $n_{\text{SiO}_2}$ $<$ $n_{\text{LC}}$. Such waveguide mode can guide the RL laterally along the metasurface, carrying the photons away from the pump location. The infiltration of NLC can considerably modify the metasurface’s supported modes, playing an important role in how directional RL can be generated.

\begin{figure*}[!htp]
    \centering
    \includegraphics[width=0.8\linewidth]{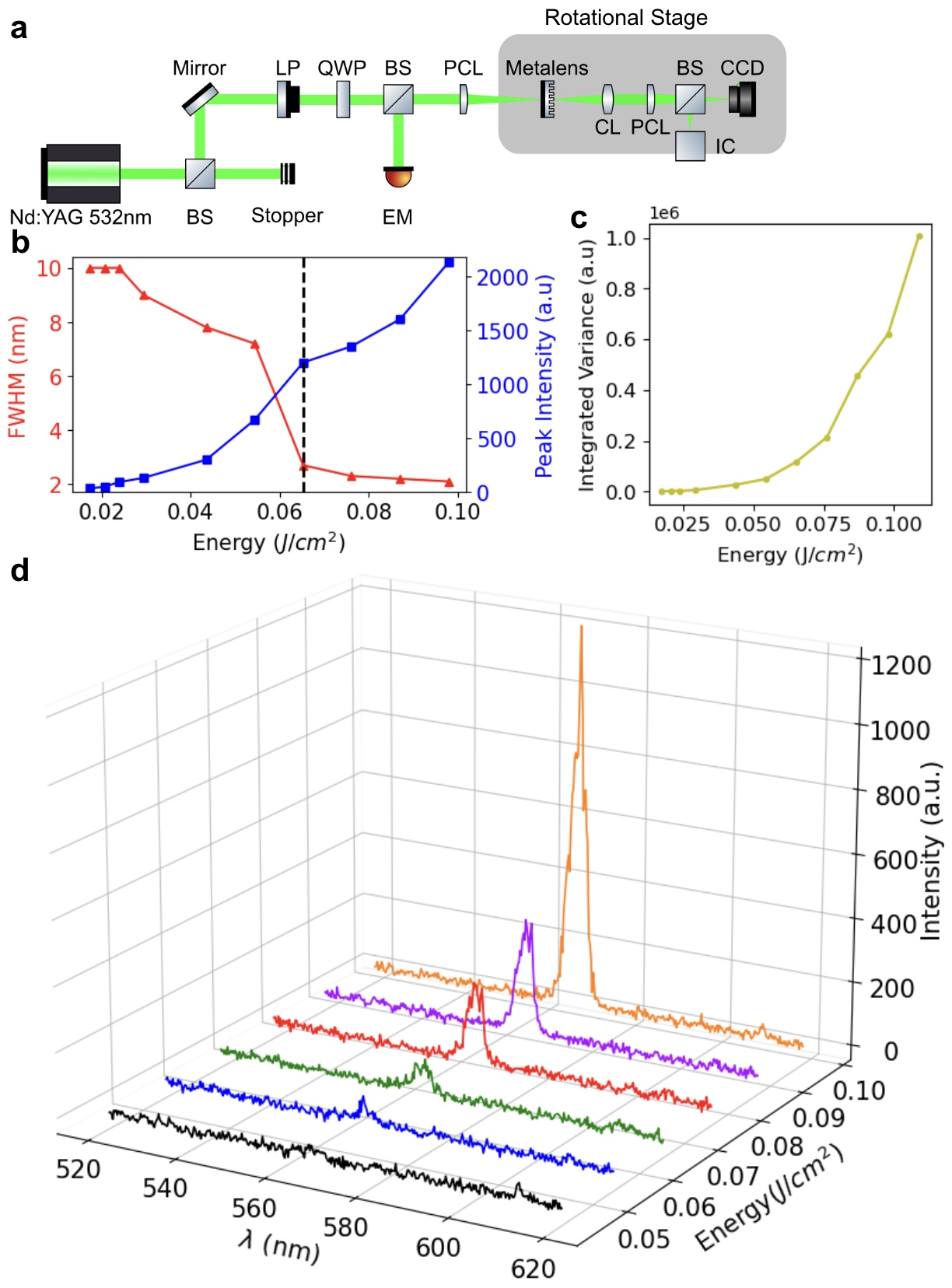}
    \caption{\textbf{ASE and RL temporal characteristics and threshold behavior.} (a) The random-lasing setup, including the metasurface and spectrometer, was mounted on a rotation stage to measure the angular emission profile. BS: beam splitter, LP: linear polarizer, QWP: quarter wave plate, PCL: planar convex lens, IC: integrating cube. (b) Lasing characteristics of PM567 in LC infiltrated metasurfaces with threshold behavior. (c) Integrated variance at various pump energies calculated over the spectrum and averages over multiple pump shots. (d) Emission spectra versus pump energy.}
    \label{fig:Fig3}
\end{figure*}

Infiltration of NLC in the metasurface is complex (Fig.\ref{fig:Fig2}b-d), depending on the hemiwicking process. The infiltration varies vastly based on geometry and material chemistry of the pillars with the NLC \cite{Lininger2020}\cite{Chen2019}\cite{Palermo2022}. By putting a small drop of NLC on the metasurface, the empty rings close to the droplet edge started pulling in the NLC through capillary force which then flows azimuthally around the lens (Fig.\ref{fig:Fig2}f,i). The NLC would flow around the ring before spreading to the next ring radially due to a more favorable path along the wider gap rings than through the narrow spacing between the pillars. The infiltration process and infiltration profile calculation are discussed further in Methods. The fully infiltrated metasurface is shown in Fig.\ref{fig:Fig2}e,h. Different color for the pillars region and empty region is observed due to different dye-doped NLC density. Under cross-polarization, the infiltrated metasurface shows four dark bands, called isogyres due to the conoscopic interference from the birefringence of the NLC. The Maltese cross suggests a configuration of either azimuthal or radial of the long-ended NLC rod molecules in the plane of the metasurface \cite{Dierking2025}. The isochromes ring due to interference between the extraordinary and ordinary rays of a certain phase difference can be observed through the empty gap rings. Two main colors of purple and grey are seen around the ring due to the inhomogeneity in infiltration from the spin coating. Rotating the metasurface under cross-polarization microscope resulted in slight color changes suggesting the LC orientation is mixed being both homogeneous and homeotropic. The infiltration profile in Fig.\ref{fig:Fig2}g follows a quadratic shape starting with a minimum at the center and flattening at 4 mm. Such infiltration agrees with the gap size of the metasurface’s empty rings during the phase jump at 0 or  $\pi$. The gap size decreases as you go further from the center following the rapidly changing phase. The narrower gap allows for a higher infiltration of NLC. The height flattens out at 4 mm where the change in the gap width and pillars distribution is now negligible.
\subsection*{Metasurface Mediated Random Lasing
}
The infiltrated metasurface was optically pumped with a 532 nm Nd:YAG laser (5 ns pulse width, 450 $\mathrm{\mathrm{\mu}}$m spot diameter, 1 Hz repetition rate). Circularly polarized excitation was used to maximize emission from the metasurface, accounting for the random molecular orientation of the LC (Fig.\ref{fig:Fig3}a). However, the polarization of the emitted light from the LC random laser does not necessarily depend on the pump polarization. The dye dipoles tend to align parallel to the LC molecules longitudinal direction which control the output polarization \cite{Yao2013}. The pump beam was positioned at different radial locations on the metasurface to probe regions with varying phase profiles. The emission from the sample is collected by a collimating lens coupled into an optical fiber connected to a spectrometer. Due to the much lower volume of dye-doped LC limited to the height of the nanopillars and subwavelength spacing of the pillars compared to conventional LC cells, the constraint on energy threshold is much higher. The slower repetition rate avoids overheating and evaporation of the LC, which can cause loss of RL due to decreasing scatterer density.

\begin{figure*}[!t]
    \centering
    \includegraphics[width=0.99\linewidth]{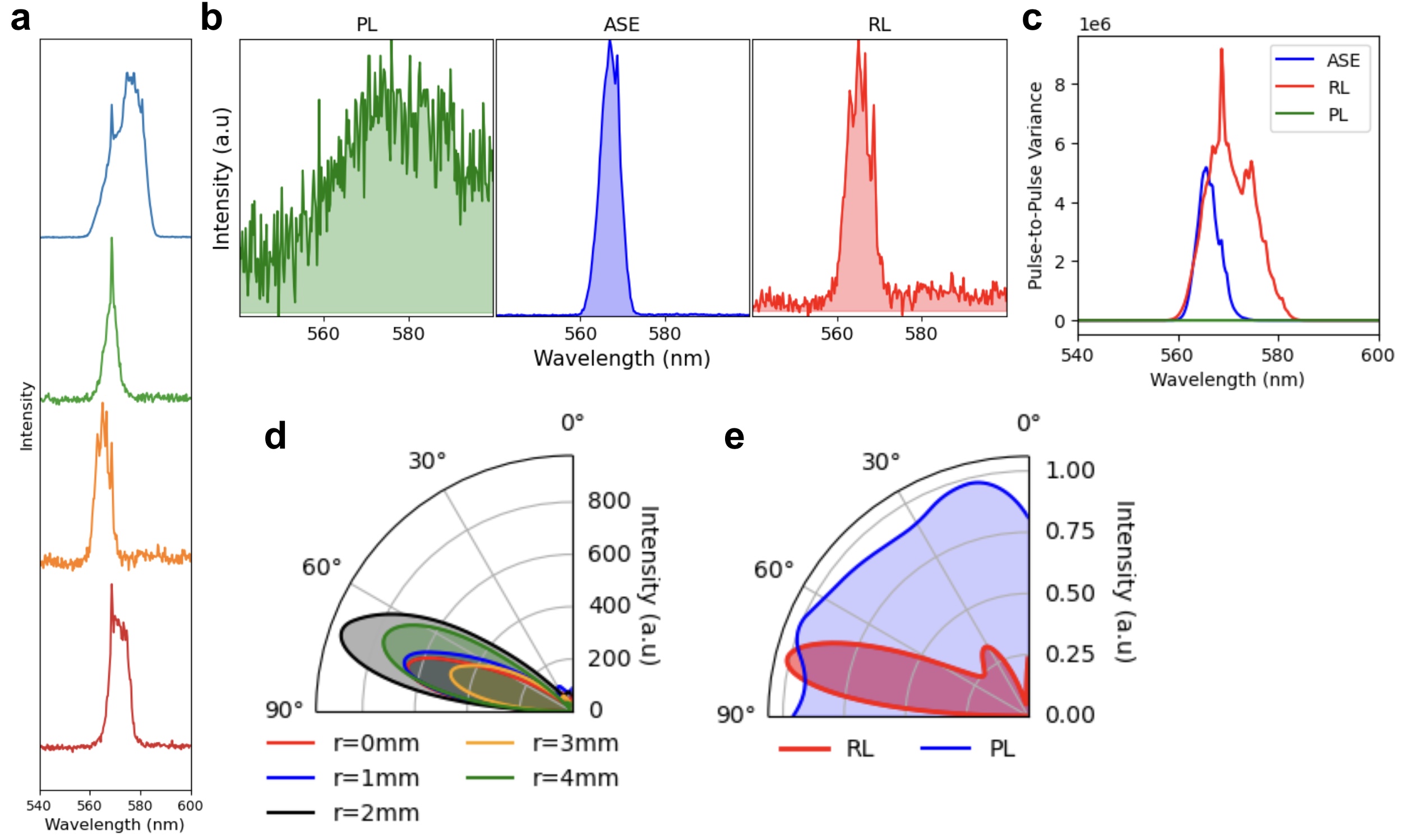}
    \caption{\textbf{Temporal and angular characteristics of random lasing.} (a) Pulse-to-pulse variation. (b) ASE, RL, and PL spectra. (c) Spectral variance. (d) Angular emission profiles of RL at different metasurface radii. (e) Comparison of angular emission for RL and PL}
    \label{fig:Fig4}
\end{figure*}

We scan the pump energy and measure the peak intensity and full-width half maximum (FWHM) of the sample emission (Fig.\ref{fig:Fig3}b). A transition from sub-threshold PL to ASE and RL above the threshold energy, E$_{\text{th}}$ = 0.065 J/cm$^2$, was observed, as evidenced by a sharp decrease in the spectral FWHM. The smaller FWHM below threshold primarily reflects the restricted emission bandwidth of spontaneous fluorescence under low excitation, further constrained by the small active volume. At low excitation energies, the combination of small pump area and limited active volume restricts emission to photons near the pump wavelength, suppressing the typical broad PL spectrum. To quantify the pulse-to-pulse characteristics of ASE, we calculated the variance at each wavelength, integrating it across the spectrum at each pump energy. The variance increased exponentially in Fig.\ref{fig:Fig3}c matching with the threshold behavior from the intensity and FWHM in Fig.\ref{fig:Fig3}b. In Fig.\ref{fig:Fig3}d, we plot the emission as a function of pump’s fluence as it transitions from the fluorescence to ASE, and RL regime. At E = 0.1 J/cm$^2$, we see a sharp lasing peak above the fluorescence curve. The fluorescence component remains visible because the limited pillar height restricts multiple scattering in the direction normal to the metasurface. Consequently, photons escape the LC more easily than those undergoing closed-loop scattering. 

\subsection*{Directional Random Lasing in Metasurface}
The system is then pumped to 1 J/cm$^2$ pushing it far beyond the RL regime. In Fig.\ref{fig:Fig4}a, the emission exhibits pronounced pulse-to-pulse fluctuations. At higher pump fluences, the fluorescence peak red-shifts due to increased reabsorption probability. Furthermore, the red-shifted fluorescence decays down faster than the ASE and RL centered around the lasing wavelength. The RL appears to have a longer lifetime due to the sustained multiple scattering process \cite{Ni2022}. The three regimes, PL, ASE, and RL are shown in Fig.\ref{fig:Fig4}b. The PL has a large FWHM, whereas the ASE has a smooth curve, with a small lasing kink at the top, and RL containing multiple sharp laser modes competing and fluctuating. The narrow FWHM of the emission indicates partial temporal coherence, proportional to $\lambda^2$/$\Delta$$\lambda$. We calculated the pulse-to-pulse variance as a function of the wavelength (Fig.\ref{fig:Fig4}c). The PL as expected shows no fluctuation with a flat variance compared to ASE and RL. While the ASE has a higher variance, with a slight kink at the lasing wavelength 567 nm, it is still steady compared to the RL. In the RL regime, the variation is dominated by the lasing modes rather than by fluorescence, as indicated by the sharp spectral peaks. The observed spectral fluctuations arise from the interplay of three factors: partial photon localization, mode competition, and LC dynamics. Localized feedback enhances narrow lasing peaks, while mode competition for gain leads to pulse-to-pulse variation in the dominant modes. Additionally, small reorientations of the LC director modify the scattering landscape and effective transport mean free path, further altering which modes reach threshold in each excitation pulse.

We then explore the directionality of the RL mediated by the metasurface. In Fig.\ref{fig:Fig4}d, we measure the RL angular profile as a function of pump position on the metasurface. In the RL regime, the angular spectra present a sharp directional emission in a cone angle around 75$^\circ$. Pumping at different positions along the metasurface yield similar directional emission around 75$^\circ$ despite having different phase profile and pillar profile. Each Fresnel zone contributes independently to the directional emission, indicating that the observed directionality stems from local coupling to guided modes rather than collective interference among FZs. Furthermore, the RL is generated within the metasurface between the pillars where the dye dipole source varies in height from 0 to 2 $\mathrm{\mathrm{\mu}}$m. The positional variation provides a different phase delay at each height, reducing the role of individual pillars as sole directional guiding factor. In addition, the infiltrated NLC layer forms a weak planar waveguide, with air and the SiO2 substrate acting as cladding and the LC as the high-index core ($\Delta$n $\approx$ 0.2). Photons generated locally from the pump location can propagate away to neighboring pillars region with momentum allowed by the waveguide mode. The dipoles within the infiltrated LC couple into the collective guided mode of the metasurface–LC structure rather than individual pillars. This mode propagates laterally supported by the waveguide mode within the metasurface and is out-coupled into air at large angles determined by the diffractive characteristics of the system as grating periodicity and refractive-index contrast. Due to the partial temporal coherence of RL, the nonlocal photons can still be outcoupled at different FZs all contributing to the high directionality. The invariance of the angular emission with pump position confirms that the observed directionality arises from the metasurface-supported mode, not from isolated pillar waveguiding. In contrast, the intensity of the spontaneous emission (PL) is nearly uniform across the quarter full rotation (Fig.\ref{fig:Fig4}e). At low pump fluence, the emission coherence is comparable to that of an incoherent source, and no directional emission is observed. In this regime, the metasurface does not influence propagation, as the emission lacks sufficient phase correlation for interference effects to occur. At low excitation energies, photons undergo multiple diffusive scattering events within the LC layer ($\ell_t$ $\gg$ $\lambda$), leading to isotropic emission. As the pump fluence increases, stronger scattering and gain localization enhance spatial coherence, reducing the transport mean free path (k$\ell_t$ $\ll$ 1) and enabling coupling to guided modes of the LC–metasurface structure. These modes outcouple into air at large angles (~75$^\circ$) through the metasurface grating, resulting in the observed directional random lasing. Although originally designed for focusing, the metasurface thus serves as a coherence-dependent outc-oupler for random lasing emission \cite{Ferjani2008}\cite{Infusino2014}.

\begin{figure*}[!t]
    \centering
    \includegraphics[width=0.99\linewidth]{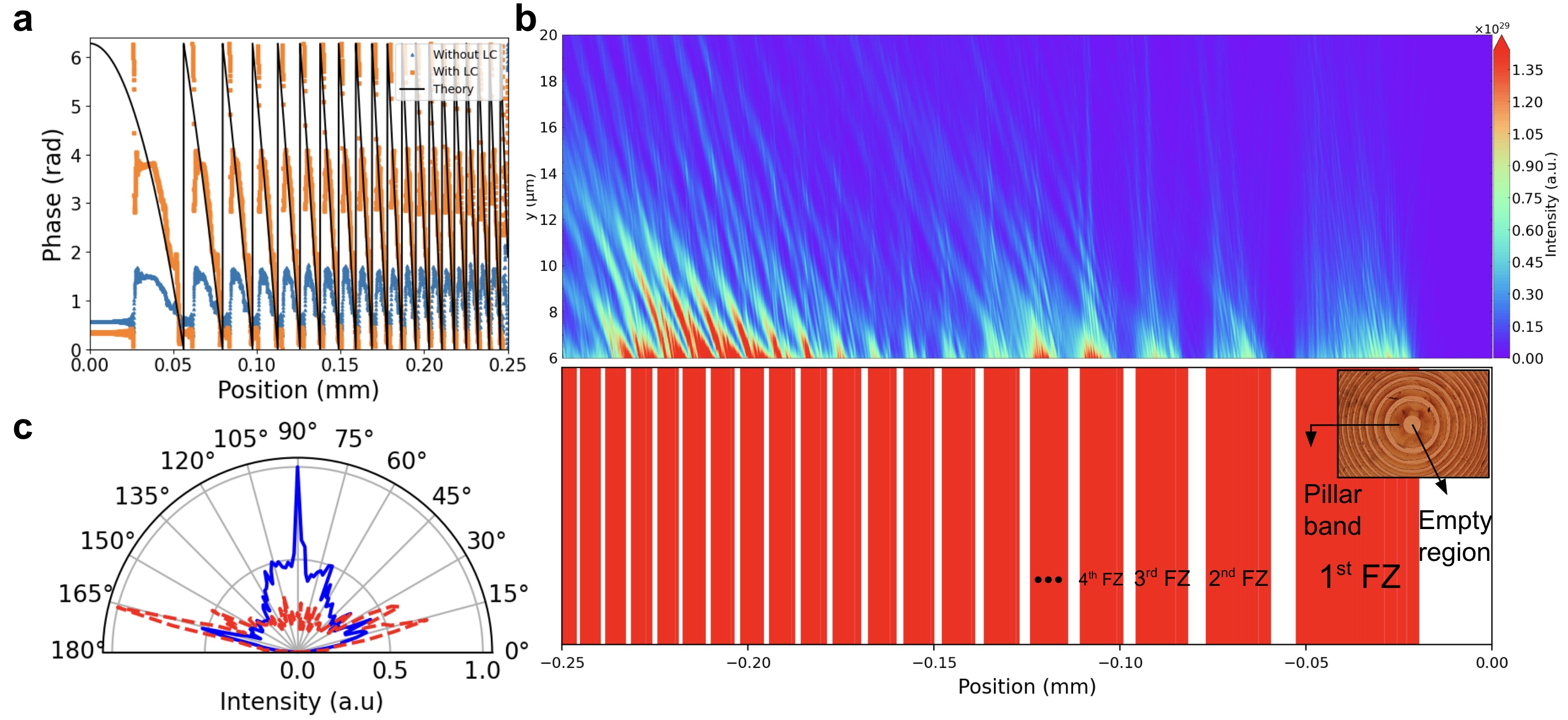}
    \caption{\textbf{Numerical modeling of metasurface-assisted emission.} (a) Target and simulated phase profiles for metsurface operation, with and without LC infiltration. (b) Far-field radiation from y-oriented dipoles (top) and metasurface structure showing FZ patterning (bottom) (inset infiltrated metasurface with FZ zones and empty rings). (c) Angular emission spectra for dipole ensembles with different orientational order parameters (S = 0.9 and 0).
}
    \label{fig:Fig5}
\end{figure*}

\textbf{Finite Element Method (FEM) simulations} were performed on the system by accounting for ensembles of dipoles within a metasurface in 2D. Details of the metasurface design and simulation procedure are provided in the Methods section. The spatial phase profile of the designed metasurface is shown in Fig.\ref{fig:Fig5}a. The phase of the metasurface is reduced due to the limited library use (from fabrication limitation) and is further reduced due to the infiltration of LC. 

In Fig.\ref{fig:Fig5}b, the simulated field from 10$\lambda$-35$\lambda$ from the metasurface is considered for one side of the metasurface, with the dipoles ensemble placed halfway at (1/2)R relative to the center where R is the radius of the metasurface. Because of the waveguiding effect introduced by the infiltrated LC, vertically oriented dipoles radiate predominantly along directions parallel to the metasurface plane. As the guided light propagates laterally, it is periodically outcoupled through the nanopillar grating, leading to interference and partial localization within each FZ. In an extended structure, the guided power would gradually attenuate or leak through the pillars before reaching the outer edge of the metasurface. We also note that the pillars in FZs across the metasurface steer the light at different angles, with higher diffracted angle coming from smaller FZs at the edge of the metasurface. This behavior is consistent with the metasurface phase profile: in geometric optics, rays farther from the metasurface center require larger deflection angles to focus on the optical axis. To clearly see the cone angle from the out-coupled light, we simulate the far-field angular profile in Fig.\ref{fig:Fig5}c. The red profile is our system in Fig.\ref{fig:Fig5}b whereas the blue profile is for an ensemble of dipoles that has zero ordering according to the definition in Maier-Saupe LC theory (more is discussed in Methods). In both cases, the nematic director is chosen to be along the y-axis. The decision comes from the LC molecules' preferential alignment along the surface. We see a large proportion of power emitted around 75$^\circ$ from the normal for the high-order orientation, which agrees well with our experimental observations for the RL’s directional emission. We still have a residual emission power at angles close to normal which is expected as not all the emitted light is guided by the LC, but rather emit directly through the air-LC interface. Upon considering the LC in the isotropic phase (blue profile), which now imposes a wider distribution of angles for the dipoles ensemble but centered at the angle perpendicular to the molecular director. The low-order ensemble exhibits a broader angular emission than the highly ordered case, with a distinct on-axis peak (0$^\circ$) arising from horizontally aligned dipoles radiating directly through the air interface rather than via guided modes in the metasurface. A weaker feature persists near 75$^\circ$, originating from a small fraction of light that couples into the nanopillar-guided mode.
\section*{Conclusion}
\vspace{-0.5em}
We have demonstrated random lasing from a dye-doped NLC infiltrated into a dielectric metasurface, where the NLC simultaneously provides optical gain and scattering. The system exhibits wide-angle directional emission under partially coherent conditions, transitioning to uniform angular photoluminescence below threshold. This simple platform requires no permanent modification of the metasurface—originally designed for focusing in free space—yet supports coherence-dependent beam steering governed by the metasurface phase geometry. The wide-angle emission remains uniform across different pump positions, confirming that the directionality arises from collective coupling within the metasurface rather than local variations in pillar width. Further tunability can be achieved by engineering metasurfaces that exploit the degree of coherence of the internal light source, enabling beam steering at specific angles and excitation conditions. Additional control is offered by the LC alignment, which can be modified through external voltage, surface rubbing, or phase transitions to tailor the scattering dynamics. Random lasing efficiency may be further enhanced through integration of plasmonic nanoparticles, which increase the local density of states, strengthen field confinement, and reduce the lasing threshold while serving as additional scattering centers.
\section*{Methods}
\subsection*{LC Preparation and Infiltration}
We are using N-(4-Methoxybenzylidene)-4- butylaniline (MBBA) Sigma Aldrich NLC which is embedded with Pyrromethene 567 (PM 567) Luxottica Exciton laser dye which has an absorption peak at 518 nm, maximum fluorescence at 547 nm and a lasing peak at 567 nm. We created a 1 wt\% of PM 567 in MBBA which we tested to be the optimized concentration to general RL without quenching at low volume and creating aggregation when infiltrating. The infiltration process of the LC inside the metasurface is complex as the non-uniform nanostructured pillars affect the wetting process. In a traditional case, when putting a liquid droplet on top of a nanostructure, we can be in two states called the Cassier-Baxter and Wenzel states. In the Cassier-Baxter state, there is no infiltration of the liquid into the air gaps between the pillars, and the droplet will stay on top of the structures. The Wenzel states only wet the airgaps which are below the surface of the liquid droplet. Both states prevent the full wetting of the nanostructures which becomes more difficult given the more intricate design of the metasurface. A distinct wetting regime, known as the hemiwicking state, occurs when a liquid droplet placed at the center of the nanostructure spreads spontaneously, forming a thin film that extends beyond the droplet’s contact area. Wetting of the nanostructured surface is governed by thermodynamic minimization of the system’s free energy, which depends on the liquid properties, pillar geometry and periodicity, and surface chemistry. To achieve uniform infiltration, a small droplet of dye-doped LC was heated to 70 $^\circ$C—transitioning to the isotropic phase—and deposited onto the metasurface. The metasurface is then held vertically and rotated for gravity to pull and spread the LC evenly across the metasurface. We then blow compressed air and put the sample on a spin coater to remove the excess solution.

To characterize the thickness of the LC, we put the infiltrated metasurface in between a polarizer-analyzer and measure the cross-intensity (co-intensity) which allows us to calculate the retardance introduced by the LC. Using the known birefringence, and the calculated retardance, we can calculate the thickness using d = $\Gamma$$\lambda$/2$\pi$$\Delta$n where $\Gamma$ is the retardance, $\lambda$ is the wavelength and $\Delta$n is the LC birefringence. The uncertainty in the infiltration profile comes from the unknown LC orientation when infiltrated inside a metasurface while having high doping concentration and subjected to the spin-coating process. We also naively assume the pre-title angle here is small which is generally not true given that there is no rubbing on the metasurface to force a certain alignment, thus allowing both in-plane and out-of-plane orientation of LC. Also, the in-plane LC may not just align along one direction having a director in a LC cell, due to the circular symmetry of the metasurface. To compensate for such an assumption, we rotate the metasurface until the largest light intensity is given for the cross-polarization state.
\subsection*{Optical Simulation}
The metasurface simulation was done using FEM COMSOL Multiphysics RF module. Due to computational power constraints, simulation was simulated in 2D with a reduced focal length of $f$ = 2.5mm but fixed numerical aperture NA = 0.1, height $h$ = 2 $\mathrm{\mathrm{\mu}}$m and edge-to-edge distance of 250 nm. A SiO$_2$ pillar library with $n$ = 1.45, diameter ranging from $d$ = 250-600 nm is employed but still maintains a 0-$\pi$ phase coverage. A flat LC layer with $n$ = 1.65 at $\lambda$ = 567 nm is assumed for simplicity with height $h_{\text{LC}}$ = 1.8 $\mathrm{\mathrm{\mu}}$m. An ensemble of electric point dipoles, N = 1000 was used to simulate the random laser. We conducted a Monte Carlo simulation averaging over 100 iterations where the position of each dipole is sampled uniformly in a specified region representing the incidence pulse width, a randomly sampled phase offset, and the orientation of the dipoles are sampled from the Maier-Saupe mean-field theory with the order parameter of the LC as a parameter. The dye molecules included in the LC tends to orient along the LC molecular director providing the ensemble a long-range order. The emission from the dipoles are collected at 2$\lambda$ above the metasurface, using Stratton-Chu near-to-far field transformation to get the angular profile in the far field. The intensities for each iteration are averaged.
\vspace{-1.5em}
\subsubsection*{Acknowledgements}
\vspace{-1.5em}
The authors acknowledge support from the Ohio
Third Frontier Project “Research Cluster on Surfaces in Advanced Materials” at Case Western Reserve University.

\vspace{-1.5em}
\subsubsection*{Data Availability}
\vspace{-1.5em}
Data are available upon request from the corresponding author. 
\vspace{-1.5em}
\subsubsection*{Ethics Declarations}
\vspace{-1.5em}
The authors declare no competing interests.

\bibliographystyle{apsrev} 
\bibliography{ref}

\end{document}